\documentclass{article}

\usepackage{url}
\usepackage{xspace}
\usepackage{booktabs}
\usepackage{natbib}
\usepackage{graphicx}

\makeatletter
\newcommand*\nobreakhyphen{\hbox{-}\nobreak\hskip\z@skip}
\makeatother

\newcommand{\up}{UniProtKB\xspace}

\newcommand{\swp}{Swiss-Prot\xspace}
\newcommand{\tr}{TrEMBL\xspace}
\newcommand{\nprot}{neXtProt\xspace}

\newcommand{\pro}{PROSITE\xspace}
\newcommand{\inter}{InterPro\xspace}
\newcommand{\pri}{PRINTS\xspace}
\newcommand{\tig}{TIGRFAMs\xspace}

\newcommand{\annotation}{textual annotation\xspace}

\title{On Patterns and Re-Use in Bioinformatics
  Databases}

\author{Michael J Bell, and
  Phillip Lord,\\
  School of Computing Science\\
  Newcastle University}

\begin{document}
\maketitle

\begin{abstract}
  
\section{Motivation:}
\label{sec:motivation}

As the quantity of data being depositing into biological databases
continues to increase, it becomes ever more vital to develop methods
that enable us to understand this data and ensure that the knowledge
is correct. It is widely-held that data percolates between different
databases, which causes particular concerns for data correctness; if
this percolation occurs, incorrect data in one database may eventually
affect many others while, conversely, corrections in one database may
fail to percolate to others.

In this paper, we test this widely-held belief by directly looking for
sentence reuse both within and between databases. Further, we
investigate patterns of how sentences are reused over time. Finally, we
consider the limitations of this form of analysis and the implications that
this may have for bioinformatics database design.

\section{Results:}

We show that reuse of annotation is common within many different databases,
and that also there is a detectable level of reuse between databases. In
addition, we show that there are patterns of reuse that have previously been
shown to be associated with percolation errors.

\end{abstract}

\section{Introduction}
\label{sec:introduction}

It is estimated that over 1,500 active databases are currently in
existence~\citep{Suarez13NAR}. While these are generally thought of as
containing biological data, they often also contain collected and collated
information about the data they carry, which is described as
\textit{annotation}. There are many different types of
annotation~\citep{Wooley05Catalyzing}: some is highly structured and organised
containing, for instance, links through to other databases, ontology terms, or
taxonomic relationships; others include unstructured or semi-structured free
text. The free text, or \textit{\annotation}, is often considered to be the
highest value annotation, although, by its very nature it is also
the hardest to represent and analyse computationally~\citep{prcis01}. In this
paper, we will consider this form of annotation.

Biological databases and annotation mirror the evolution of biological
systems. As highly similar genes occur in many different organisms, transferred
both horizontally and vertically, so the annotation about these genes is
reused between different databases~\citep{Richardson01012013}. This form of
reuse substantially reduces the work required by database annotators, but also
creates a problem; for most databases it is difficult to determine the source
or support for a particular statement~\citep{DBLP:journals/corr/abs-1012-1660}.
Although most databases contain out-going references, either to the primary
literature or to other databases, these are generally given at the level of
the database record. For the richest databases, there may be many statements
for each record. In short, databases often lack a formal representation of
their \textit{provenance}~\citep{DBLP:journals/corr/abs-1012-1660}.

While this form of reuse is part of the folk-history of
bioinformatics\footnote{\small\url{http://madhadron.com/posts/2012-03-26-a-farewell-to-bioinformatics.html}},
and is apparent from even a short perusal of a few bioinformatics
databases, it has rarely been explicitly studied. In a previous
study~\citep{bell2013can}, we have shown that the level of reuse in
\up is extremely high -- the most reused sentence in \tr occurs more
than seven million times, while the most common sentence in \swp
occurs more than 91,000 times. More over, we have shown that this
reuse operates as an informal indicator of provenance; two identical
sentences are likely to share a common history. This, in turn, allowed
us to identify \emph{propagation patterns} that can be used to detect
inconsistencies and errors in this annotation.



Our previous analysis looked at only a single database; but we also believe
that reuse occurs between different databases, forming a biological knowledge
ecosystem. In this paper, we extend the analysis further, looking at several
different databases, and show that \textit{within} these there are also high
levels of reuse. The same analysis also allows us to track reuse
\textit{between} databases and show that, here also, there is significant reuse.
Further, we look for tell-tale signature patterns previously shown to indicate
erroneous annotation and show that these patterns are also present within several
databases and can be seen between several databases. This analysis suggests
that as well as reuse being common-place, that it is possible to detect
knowledge flow between databases, giving an informal mechanism for detection
of provenance.

\section{Methods}
\label{sec:methods}

\subsection{Choosing a set of Databases}



There are many databases in the bioinformatics ecosystem that we could
use to study. Unfortunately, these vary significantly technologically,
both in terms of their format, their identifiers and their scheme for
updates and maintenance history. Our previous analysis focused
exclusively on \up, using it an exemplar gold standard. This analysis
also benefited from the organisation of \up, which consists of two
databases: \swp, which is manually curated and reviewed; and \tr,
which is computationally generated and unreviewed. Here, we wish to
identify a set of suitable databases that allow us to extend our
analysis further.

We, therefore, have used the following criteria for selection of a database,
firstly on technical grounds: the database must make available historical
versions; contain more than just minimal amounts of \annotation; and, be in a
form which is relatively easy to obtain and parse. Within this, we have picked
a set of databases of mixed maturity to obtain a reasonable sample. We chose
the following five databases:

\begin{itemize}
\item \nprot~\citep{Lane12neXtProt} --- focused solely on human proteins,
  \nprot incorporates data from various sources and is built as a
  participative platform; the core corpus is based on human proteins
  from \swp. Unlike many databases, \nprot provides a classification
  system that categorises data based on its quality into gold, silver
  or bronze. 

\item \pro~\citep{sigrist2013new} --- consists of sequence patterns, or
  motifs, that are conserved in protein sequences and can be used to
  help infer information about a sequence, such as which protein
  family it belongs to and its possible function.  Each \pro entry
  contains a pointer to a relevant documentation entry, which provides
  biological information that can be inferred by the pattern. 

\item \pri~\citep{attwood2003prints} --- a collection of sequence
  motifs, similar to \pro. However, entries in \pri are known as
  fingerprints, as they are composed of multiple motifs, unlike
  entries in \pro which contain only single motifs. All \pri entries
  are manually curated and provide cross-references to the equivalent
  \pro entries, if they exist.

\item \tig~\citep{haft2003tigrfams} --- provides a collection of protein
  families which are designed to assist with the prediction of protein
  function. Each \tig entry contains a \annotation section with
  additional supporting information, such as GO annotations and
  references to relevant Pfam and \inter entries.

\item \inter~\citep{hunter2012interpro} --- an integrative database
  collating information regarding protein families, domains and
  functional sites from eleven member databases, including \pro, \pri
  and \tig. Each \inter entry contains a description, or abstract,
  which is often supplemented with references to relevant literature.
\end{itemize} 

The chosen databases are summarised, along with the URL used to access
each database, in Table~\ref{tab:databaseURLs}.

\begin{table*}[!ht]
  \footnotesize \centering
  \begin{tabular}{lll}
  \toprule
  Database Name & URL \\
  \midrule
  \up (\swp \& \tr) & \url{http://www.uniprot.org/} \\
  \inter & \url{http://www.ebi.ac.uk/interpro/} \\
  \nprot & \url{http://www.nextprot.org} \\
  \pro & \url{http://prosite.expasy.org/} \\
  \pri & \url{http://130.88.97.239/PRINTS/} \\
  \tig & \url{http://www.jcvi.org/cgi-bin/tigrfams/} \\
  \bottomrule
\end{tabular}
\caption{The databases chosen for our analyses, including the web address (URL) of each database.}
\label{tab:databaseURLs}
\end{table*}

\subsection{Data extraction and visualisation}

For our analysis, we need to extract sentences from the \annotation of
each database. As each of these has a different format for each of
these necessitates, a custom framework was written for each, which was
extended from the tool described
previously~\citep{bell2013can}. Fortunately, the requirements for our
analysis are fairly simple: we need only extract the \annotation and
basic metadata for a record (the identifier or accession number), so
this process is relatively straightforward and robust to differences
(or changes over time) in the database format. Sentences are
intentionally extracted \emph{verbatim} and stored in lower-case, with
only database-specific formatting removed. For example, the following
data from \up:

\begin{tiny}
\begin{verbatim}
CC -!- FUNCTION: May be a transcription factor with important functions
CC     in eye and nasal development.
\end{verbatim}
\end{tiny}

would be transformed and stored as:

\begin{tiny}
\begin{verbatim}
may be a transcription factor with important functions in eye and nasal
development.
\end{verbatim}
\end{tiny}

This form of analysis is intentionally very simple; we performed no stemming
or even stop-word analysis, with white space normalisation the only change
made to sentences. While this form of analysis may seem very blunt, we choose
it for two reasons: it is computationally very attractive, both when parsing
and searching for matches; and, most importantly, we were concerned more with
correctness than recall. When a match between two databases is found, it is
very likely to be a real one.

Following extraction, sentences were stored in a relational database, linked to
a record identifier, database and version. Dates of records are calculated
using the release version in which a record occurs, and therefore reflect an
upper bound, the size of which is reflective of the release frequencies of the
databases, as described previously~\citep{bell2013can}.

The visualisation of sentence propagation uses an interactive
visualisation using the Highcharts library~\footnote{\url{http://www.highcharts.com/products/highcharts}} driven
directly from the database generated in the previous step. These
visualisations provide various interactive features such as zooming,
narrowing and so forth. For full details, please
see~\citep{bell15:provenance_and_propagation}.

\section{Results}
\label{sec:results}

First, we introduce a number of measures that we have used to analyse
reuse of \annotation. We focus on the number of
sentences within a database. It would be expected that for the
sentences that occur in the database, some will occur more than once
(i.e.\ the database is redundant) and some only once. These allows us
to distinguish between the three following measures of a sentence
which we reuse throughout the paper.

\begin{itemize}
\item Total sentences -- A redundant set of all sentences in a
  database version.
\item Unique sentences -- A non-redundant set of all sentences in a
  database version.
\item Singleton sentences -- A set of sentences that occur only a
  single time within an entire database version.
\end{itemize}

\subsection{Reuse within Databases}
\label{sec:reuse-with-datab}

Previously, we have shown that \up (i.e. \swp and \tr) reuse sentences between
multiple records; in the case of \tr this reuse is extreme with only 8,131
unique sentences from 22,706,421 total sentences. First, we address the
question of how widespread this practice of reuse is within our chosen
databases. Moreover, we ask whether this is a feature of the overall size and
complexity of a database.

\begin{figure}[!tpb]
\centerline{\includegraphics[scale=0.2]{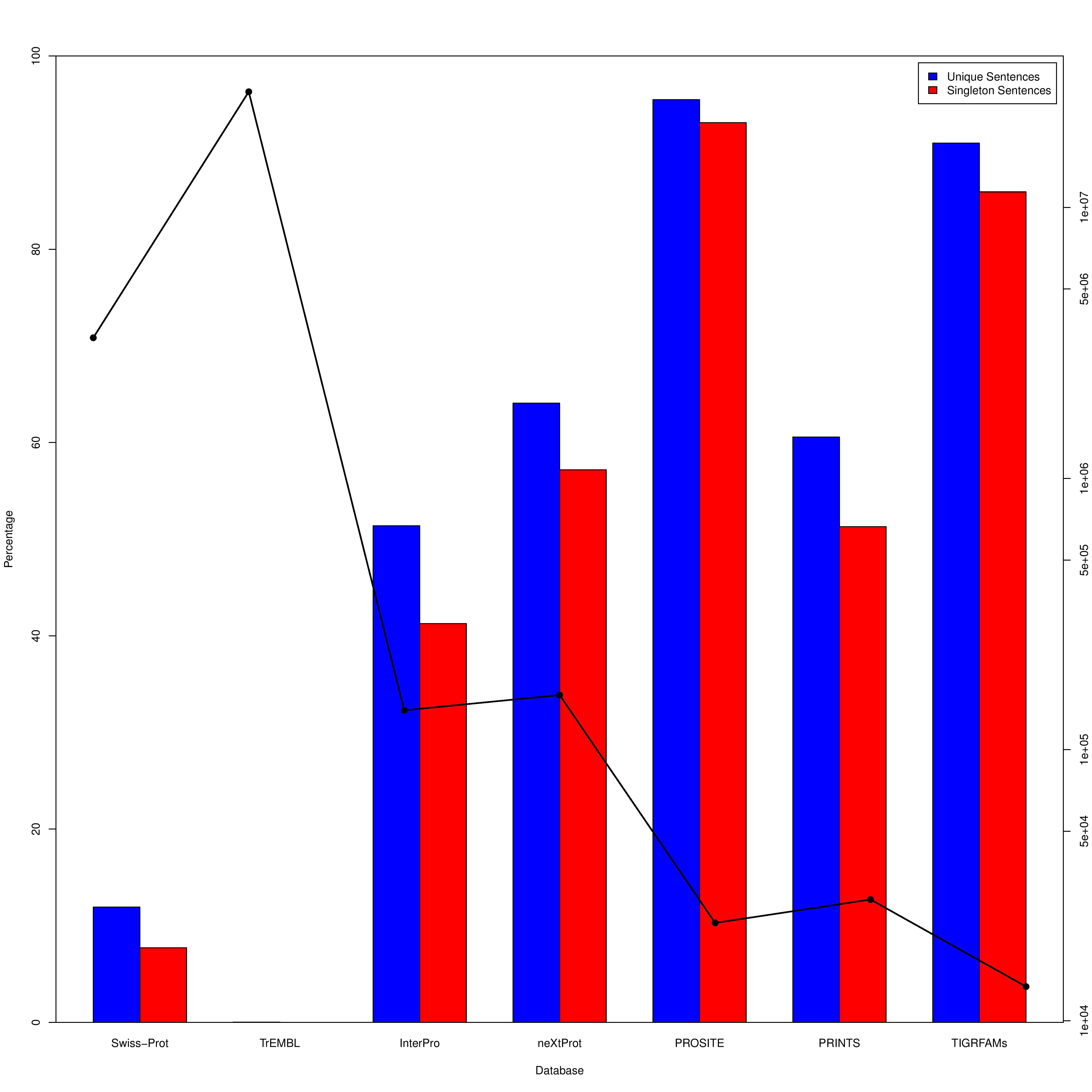}}
\caption{Figure showing the percentage of singleton (red) and unique
  (blue) sentences in each analysed database. The line graph
  represents the total number of sentences in the database (shown on
  log scales). Within this graph we can broadly see that the larger
  the database, the more redundant its annotation.}
\label{fig:totalAndPercentSentences}
\end{figure}

To address this question, in Figure~\ref{fig:totalAndPercentSentences}
we show the total number of sentences in each database; to recap, this
is the number of sentences that occur in all records, whether they are
duplicates or not. This is shown on a log scale as \tr is much larger
in size than all of the others, as shown in
Table~\ref{tab:sentenceDatabaseSummary}. In
Figure~\ref{fig:totalAndPercentSentences}, we also show the number of
unique and singleton sentences as a percentage of the total.

\begin{table*}[!ht]
  \footnotesize
  \centering
  \begin{tabular}{lllll}
  \toprule & Total Sentences & Unique Sentences &
  Singleton Sentences & Total Unique \\
  \midrule
  \swp & $3,304,681$ & $394,233$ & $255,349$ & $531,206$ \\ 
  \tr & $26,706,421$ & $8,131$ & $735$ & $49,665$  \\ 
  \inter & $139,624$ & $71,755$ & $57,628$ & $100,874$ \\ 
  \nprot & $158,929$ & $101,822$ & $90,875$ & $110,607$ \\ 
  \pro & $22,940$ & $21,902$ & $21,356$ & $29,127$  \\ 
  \pri & $27,987$ & $16,953$ & $14,356$ & $17,858$  \\ 
  \tig & $13,360$ & $12,155$ & $11,481$ & $13,373$ \\  
  \bottomrule
\end{tabular}
\caption{Table showing the total number of sentences, unique (i.e.\
  distinct) sentences and singleton sentences contained within the
  latest version of each analysed database. Additionally, we show the
  total number of unique sentences over the lifetime of the entire
  database.} 
\label{tab:sentenceDatabaseSummary}
\end{table*}

This analysis shows a number of features. First, \tr is shown to be an
extreme outlier; as a database, it is very large, but has the lowest
number of unique sentences (8,131, followed by \tig with 12,155). Of
the databases, the \pro database has the highest percentage of unique
sentences -- over 95\% of sentences are unique. Broadly, this theme is
also repeated in the other databases -- the larger the database, the
more reuse we see.

From this, we conclude that reuse of sentences is a feature of all of the
databases that we have analysed, and that this reuse is substantial in most
cases. 

\subsection{Patterns of Reuse within databases}
\label{sec:patt-reuse-with}

Previously, we have shown that there are identifiable
\textit{patterns} of reuse within \up. We hypothesised that some of
these patterns may be indicative of low quality or erroneous
annotation occurring as a result of a failure to propagate changes;
this was confirmed for one pattern by a close analysis of a number of
examples~\cite{bell2013can}.

Having confirmed in Section~\ref{sec:reuse-with-datab} that sentence reuse is
a feature of all databases that we have analysed. We now address the question
as to whether the patterns we found in \up are also present elsewhere.

We analyse the databases here for two patterns, \emph{transient} and
\emph{missing origin}. The transient pattern is where sentences occur within
an entry for only a single database release (i.e.\ they are removed from an
entry after one iteration of the database). From this definition, it follows
that it is impossible to classify a sentence as transient when it occurs only
in the current version of a database, so we show these independently as
\emph{possibly transient}, although we do not consider this to be a separate
pattern. A sentence follows the missing origin pattern if it initially occurs
in a database entry, is later propagated to a secondary entry (or entries) and
then subsequently removed from the origin entry whilst still remaining in the
secondary entries. Table~\ref{tab:missingOriginAllDatabases} shows the number
of sentences identified in each database which follow each pattern.

\begin{table*}[!ht]
  \footnotesize \centering
  \begin{tabular}{llll}
  \toprule
  Database Name & Missing Origin & Transient & Possibly Transient\\ 
  \midrule
  \up & $8,355$ & $42,460$ & $25,582$\\ 
  \inter & $2,689$ & $4,094$ & $1,293$\\ 
  \nprot & $35$ & $5,148$ & $773$ \\ 
  \pro & $132$ & $2,644$ & $21$\\ 
  \pri & $81$ & $206$ & $363$\\ 
  \tig & $17$ & $563$ & $63$\\ 
  \bottomrule
\end{tabular}
\caption{Table summarising the number of sentences following the transient and missing origin
  propagation patterns for each database. Sentences classified as
  possibly transient are those which appear a single time in the
  latest version of the database.}
\label{tab:missingOriginAllDatabases}
\end{table*}

\begin{figure}[!tpb]
\centerline{\includegraphics[scale=0.42]{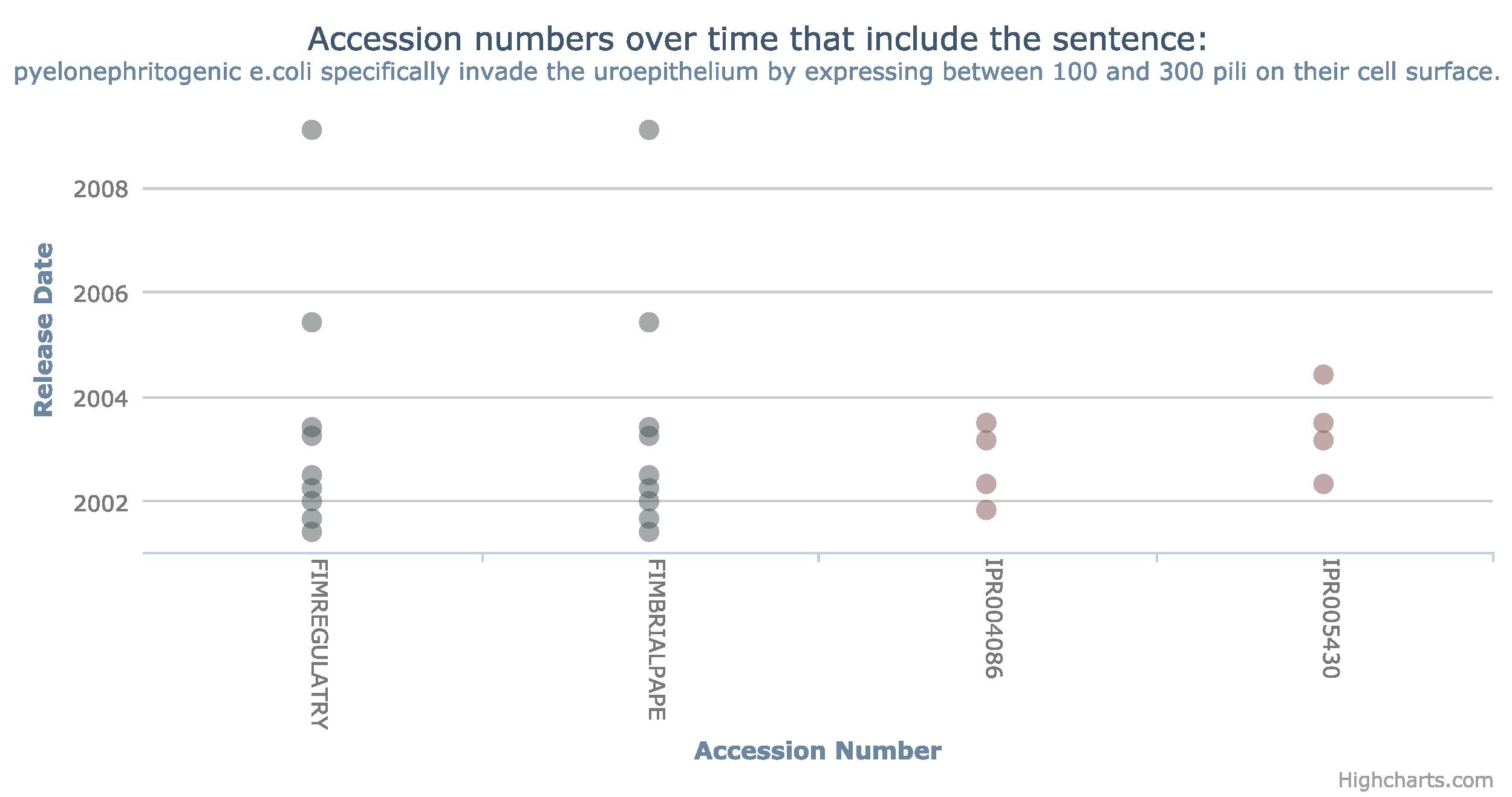}}
\caption{Example of sentence which follows the missing origin pattern. Here,
  the sentence originates in \inter entry IPR004086 before later appearing in
  entry IPR005430. It remains in this entry even when then sentence is removed
  from IPR004086. Interestingly, we note that the sentence occurs in \pri both
  before and after it exists in \inter.}
\label{fig:missingOriginInterpro}
\end{figure}

From these results, we note that all of the databases show incidences of the
patterns that we have previously identified. Of the databases, \pri and \tig
have the lowest level of all of these patterns. This is consistent with
Figure~\ref{fig:totalAndPercentSentences} -- as these patterns are a feature
of a unique sentence, they are upper-bounded by the uniqueness, and likely to
be affected by the level of reuse within the databases. To be classified, a
sentence only needs to exhibit a pattern in a single entry. A clear example of
this is shown in Figure~\ref{fig:missingOriginInterpro} which shows an example
of the missing origin pattern. This sentence (``pyelonephritogenic e.coli
specifically invade the uroepithelium by expressing between 100 and 300 pili
on their cell surface'') initially appears in \inter entry IPR004086 in 2001
and later appears in \inter entry IPR005430 approximately a year
later. However, the sentence is removed from IPR004086 (the origin) in 2003
while still remaining in the secondary entry IPR005430 for another release.

We have chosen this example, because it clearly represents an error in the
database, albeit a minor typographical one; namely the presence of a space
between the species and genus in ``E. coli''\footnote{This modification seems
  to reflect a change in the underlying XML representation as taxonomic markup
  was removed at the same time. Our analysis explicitly excludes markup early
  in the pipeline; we note, however, that were it included, the missing origin
  pattern would also have detected the lack of percolation of markup
  changes.}. Obviously this form of the error is unlikely to cause major
challenges for human consumption of the database annotation, but could cause
issues for computational use.

In this section, therefore, we have demonstrated that the patterns of reuse
that we have previously seen in \up also occur in other databases and in some
cases, at reasonably high levels. In general, these patterns occur more in
databases with more redundancy.

\subsection{Reuse between Databases}
\label{sec:reuse-betw-datab}

One common worry about knowledge in biology is that it is circular, as the
knowledge is reused and percolated through the biological database ecosystem.
If this is true, then we should be able to detect this supposed reuse, for
instance by sentence reuse \textit{between} databases. In fact
Figure~\ref{fig:missingOriginInterpro}, in addition to showing a missing
original, also shows an example of reuse between databases; a sentence which
appears first in \pri and, then, later reappears in \inter. 

To address this question more systematically, we have looked for identical
sentences that occur between any of the databases in our collection, the
results of which are shown in Table~\ref{tab:propagationSummary}.

\begin{table}[!ht]
  \footnotesize
  \centering
  \begin{tabular}{ll}
  \toprule
  Database Combination & Total Sentences \\ 
  \midrule
  \up & $526,435$ \\ 
  \nprot; \up & $83,868$ \\ 
  \inter & $82,968$ \\ 
  \nprot & $26,539$ \\ 
  \pro & $23,182$ \\ 
  \pri & $10,064$ \\ 
  \tig & $9,661$ \\ 
  \inter; \pri & $7,751$ \\ 
  \inter; \pro & $5,790$ \\ 
  \inter; \tig & $3,681$ \\ 
  \inter; \up & $435$ \\ 
  \inter; \nprot; \up & $151$ \\ 
  \pro; \up & $71$ \\ 
  \inter; \pro; \up & $26$ \\ 
  \nprot; \pro; \up & $20$ \\ 
  \inter; \pri; \up & $20$ \\ 
  \inter; \nprot; \pro; \up & $19$ \\ 
  \inter; \pri; \pro & $14$ \\ 
  \tig; \up & $14$ \\ 
  \inter; \tig; \up & $9$ \\ 
  \inter; \nprot; \pri; \up & $4$ \\ 
  \nprot; \tig; \up & $3$ \\ 
  \inter; \nprot; \tig; \up & $2$ \\ 
  \inter; \tig; \pro & $2$ \\ 
  \inter; \nprot; \pri; \pro; \up & $1$ \\ 
  \inter; \pri; \pro; \up & $1$ \\ 
  \pri; \up & $1$ \\ 
  \pri; \pro & $1$ \\ 
  \pri; \tig & $1$ \\ \bottomrule
\end{tabular}
\caption{Table summarising the distribution of all unique
  sentences shared between the analysed databases.}
\label{tab:propagationSummary}
\end{table}

These results show that there is substantial reuse of sentences in two key
cases. Firstly, there is a very high-level of reuse between \up and \nprot.
This is expected as \nprot explicitly depends on \up -- in this case,
perhaps, it is more surprising that a significant proportion of \nprot is
unique to it (around 25\% of the total sentences in \nprot). A second case
is shown between the \inter database and \pri, \pro and \tig. This
is to be expected as \inter is a federated database, explicitly depending on
the other three databases. We do also see reuse between other databases,
although this occurs at a fairly low-level, compared to the total number of
sentences. There is one sentence which occurs in all five of the databases
which is \textit{``visual pigments are the light-absorbing molecules that mediate
vision.''}

From this we conclude that knowledge does percolate between different
databases and that it is possible to detect this by using whole sentence
analysis. However, in the majority of cases where identical sentences are
found in large numbers between databases, occur as a result of a formal
relationship between the two -- for instance, between \up and \nprot.

\subsection{Patterns between databases}
\label{sec:patt-betw-datab}

As we have shown previously, and in this paper, it is possible to detect
patterns of reuse within databases, and that in some cases these patterns
appear to be related to errors of percolation. Further, we know that, in some
cases, sentence percolation can also be seen between databases. This raises
the question as to whether we could detect patterns that occur between
databases.

While we do have algorithms for pattern detection within a database, the same
process turns out to be considerably harder between databases, mostly because
of the lack of co-ordinated release dates. If a database record contains a
sentence which is removed between two releases, for example, should it be
considered present only till the first release, or till just before the
second? When comparing two databases, these problems are significant, as the second
database may have undergone several releases subsequently.

\begin{figure}[!tpb]
\centerline{\includegraphics[scale=0.42]{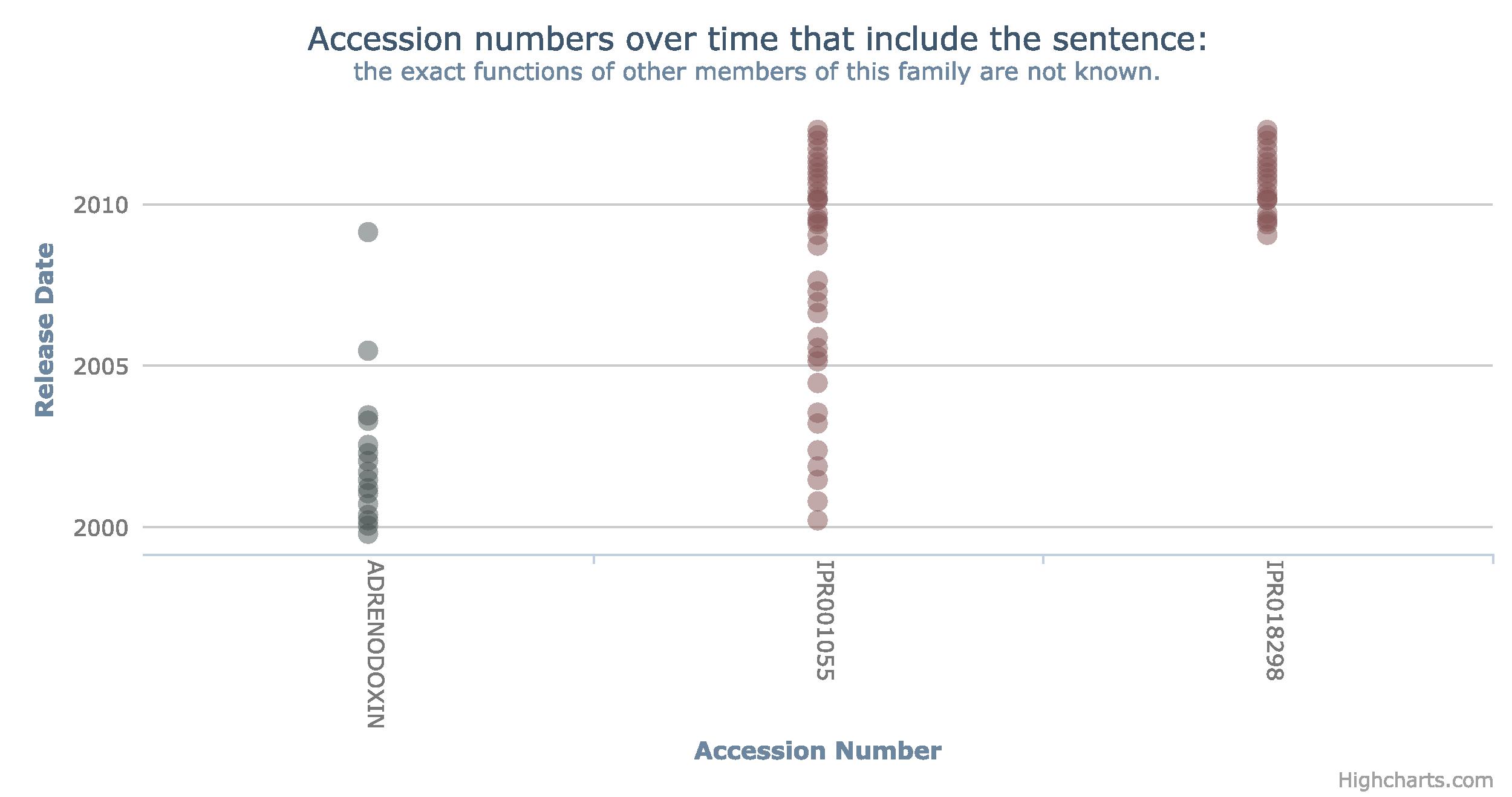}}
\caption{An example of a sentence in \inter which does not follow any
  propagation pattern. However, if you also consider \pri, and the
  sentence was copied from \pri into \inter, then the sentence
  technically follows the missing origin pattern. This would have
  significant impact on the potential correctness of sentences in all
  databases.}
\label{fig:propagationAndmissingOrigin}
\end{figure}

As a result of these issues, we have not yet been able to address the question
of pattern occurrence systematically between all databases. However, we have
been able to find specific examples by inspection. We show one of these in
Figure~\ref{fig:propagationAndmissingOrigin}. In this case, a sentence appears
first in \pri (in around 1999), and then later in 2000 appears, presumably
by percolation, in \inter first in one record (IRP001055) and then later in
2008 in another (IPR018298). Around the same time, it disappears from the
original entry.

Interestingly, it is not possible to detect the occurrence of this pattern
just by considering a single database. In \pri, the sentence occurs at one
point, then stops later. In \inter, it continues to occur in all
records that it has percolated to. It is only by considering the removal from
\pri, and the continued occurrence in \inter that we see an instance of the
missing origin pattern. This does suggest that cross-database comparisons may
reveal more knowledge than the consideration of a single database.

Of course, inspection of this form is not a scalable mechanism for detecting
instances of these patterns; however, without co-ordinated release dates,
automation is hard to achieve. Despite this, our initial analysis indicates
that there are examples of reuse patterns that are detectable between
different databases and, indeed, patterns that are only detectable by
considering multiple databases.

\section{Discussion}
\label{sec:discussion}

It is often said that knowledge in bioinformatics is frequently
reused, and moves through the database infrastructure. In this paper,
we have attempted to investigate this in as direct a manner as
possible, by looking for exact reuse of sentences between different
databases. As a result of this we have found that, indeed, reuse of
knowledge within bioinformatics databases is extremely common. In the
most extreme case for a manually curated database, \up/\swp, some 91\%
of the sentences occur more than once. For \tr, this is even more
skewed where unique sentences number less than 1\% of the total
sentences. To our knowledge, this demonstrates the first system
attempt to detect, investigate and record the impact of this knowledge
flow.

This reuse is, perhaps, a reflection of evolution of these databases. By
itself, it is not necessarily a problem, however, it is a cause for concern.
It does mean that the annotation is heavily \textit{denormalised} -- that is,
what is effectively the same data is stored multiple times within a single
database. This presents significant difficulties during updates; if a
duplicated piece of knowledge needs to be updated with respect to a single
record, then perhaps it also needs to be updated with respect to another.

We have previously shown that it is possible to detect errors or low-quality
annotation, resulting from this denormalisation, by looking for specific
patterns of provenance in the database~\citep{bell2013can}. In this paper, we
have shown that two of these patterns, the missing origin and transient
patterns, are also present within other databases; in the case of the missing
origin pattern, this is clearly in the systematic representation of species
names. Further, we have shown that reuse also occurs between databases
although, in general, this happens at a fairly low-level. Even here, though,
it is possible to detect patterns between databases where they are not
detectable from a single database.

The work described here shows the value and importance of historical records,
and that this value is also relevant to the present. We have previously made
extensive use of historical records when looking at trends in database word
usage~\citep{greycite2243}, as have others to determine when a database might
be complete~\citep{Baumgartner2007}, or to assay the accuracy of predictive
tools~\citep{Gross09Estimating}. These analyses have dealt with both the
structured (GO) and unstructured (comments) components of
annotation. This demonstrates that an accurate record of the past is useful to
increase our understanding of the current state of the annotation; truly,
understanding the past is useful to correcting the errors of the present.

However, there are important limitations. In our previous work, we were more
able to investigate some of the instances of annotation patterns in detail,
and demonstrate that they were actually errors. In this work, we were greatly
aided by the existence of UniSave~\citep{leinonen2006unisave} which allowed us
to rapidly and efficiently investigate the historical record. \up is
unusual in providing this form of tool however.

We can compare this to Wikipedia which includes a more complete feature set
with respect to versioning than any of the bioinformatics databases that we
have analysed (with \up coming a notable second best). It does
demonstrate that it is possible to store a fine-grained full version history
for even a very large knowledge base. That it is searchable using the current
schema is an added bonus and would greatly help this form of analysis; in
fact, Wikipedia has been used as the basis for analysis of historical
resources~\citep{Viegas04Studying}. Interestingly, in the last few years, PFAM
has moved toward using Wikipedia as the main mechanism for maintaining their
textual annotation~\citep{Punta01012012}; while we do not believe this was the
original intention, from the point-of-view of this analysis, this move should
increase the quality of the historical data available.

The second critical limitation of our work is that we are not looking directly
at provenance but inferring from the occurrence of identical sentences. In our
work, we have erred on the side of caution by using direct string matching;
this is a very useful tool for two reasons: firstly, it is computationally
very simple, and extremely scalable and secondly it gives a high-level of
confidence that a match does actually demonstrate knowledge flow. It is,
however, also a very blunt tool, and we are likely to be missing many examples
of information flow. Small changes to sentences, including grammatical or
textual corrections, will break the provenance trail; indeed, we have a direct
example of this happening. Moreover, when tracking provenance between
databases, we suspect that database authors have a positive incentive to alter
text to avoid issues of copyright or plagiarism, inadvertently making the
provenance even harder to track.


A third issue with tracking provenance is the difficulty of dating individual
sentences. Databases are normally developed continuously, but only released
periodically, and it is the releases that we have tracked. These problems are
exacerbated between databases, as the release date is the only information we
have to infer the direction of the travel of knowledge. Taken together, these
limitations mean that our understanding of provenance is heuristic and may be
wrong. In short, our ability to exploit this knowledge is curtailed by the
limited provenance information that is stored.

There are practical steps that current database provider could take which
could increase our knowledge. Most software engineering projects make use of
version control, which can store practically unlimited provenance of source
code. Wikipedia (and, therefore, also PFAM) use the same technology for their
textual annotation. This may provide a simple solution for many biological
databases; it would, at least, address the requirement for fine-grained date
information.  Alternatively, a more formal model of provenance (such as
PROV~\citep{Missier:2013:WPF:2452376.2452478}) might be used, which could
potentially provide a more fine-grained dataset describing the relationships
between sentences explicitly. This is also likely to be necessary for larger
databases such as \tr, which may be less suited to version control systems
because of their size, automatic generation and relatively low levels of
\annotation per entry.


Despite these limitations, we have shown that knowledge flows
between databases even when there is not a formal link between them. While this
raises the spectre that some of the knowledge in these databases may be
circular, we have also shown that it is possible to detect patterns which may
lead to mechanisms of error detection, which should increase the quality of
knowledge in biology.

\section{Acknowledgements}

MJB was funded by an EPSRC DTA award to Newcastle University.

\bibliographystyle{unsrtnat}
\bibliography{bibliography}

\end{document}